 \documentclass[aps,prc,showpacs,superscriptaddress,preprint,floatfix]{revtex4-1}

\usepackage[dvips]{graphicx}
\usepackage{color}
\usepackage{amssymb}
\usepackage{amsmath}
\usepackage{braket}
\usepackage{hyperref}
\usepackage{bm}
\usepackage{multirow}
\newcommand{\nc}{\newcommand}           % new command
\nc{\vc}[1]     {\mbox{\boldmath $#1$}} % boldmath(vector)
\nc{\mapleft}[1]{                       % something under arrow
 \smash{\mathop{                        %
  \hbox to 0.90cm{\rightarrowfill} }\limits_{#1}}}
\nc{\figwidth}{0.8}                    % figure width, in unit of textwidth, preprint
\nc{\mydraft}	{\setlength{\topmargin}{-1.5cm}}
\mydraft   

\begin{document}
\title{Microscopic investigation of one- and two-proton decay from the excited states of $^{10}$C}

\author{Qing Zhao} \email[]{zhaoqing91@zjhu.edu.cn}
\affiliation{School of Science, Huzhou University, Huzhou 313000, Zhejiang, China}
\author{Masaaki Kimura}
\affiliation{Nuclear Reaction Data Centre (JCPRG), Hokkaido University, Sapporo 060-0810, Japan}
\affiliation{Department of Physics, Hokkaido University, Sapporo 060-0810, Japan}
\affiliation{RIKEN Nishina Center, Wako, Saitama 351-0198, Japan}
\author{Bo Zhou}
\affiliation{Key Laboratory of Nuclear Physics and Ion-beam Application (MOE), Institute of Modern Physics, Fudan University, Shanghai 200433, China}
\affiliation{Shanghai Research Center for Theoretical Nuclear Physics, NSFC and Fudan University, Shanghai 200438, China}
\author{Seung-heon Shin}
\affiliation{Department of Physics, Hokkaido University, Sapporo 060-0810, Japan}

% cSpell: enable

\begin{abstract}
We present microscopic cluster model calculations for the $1p$ and $2p$ decays of the $0^+_2$ and $2^+_2$ states of $^{10}$C. With the $R$-matrix method, we have estimated the decay widths. The obtained $1p$ and $2p$ decay widths are in good agreement with the recent experimental data and support the validity of the di-proton approximation for the $2p$ decay of $^{10}$C($0^+_2$). We also show the suppression of the $1p$ decay of the $^{10}$C($0^+_2$) state due to the structure mismatch with the decay channel.
\end{abstract}

\maketitle

\section{Introduction}
The study of proton decay is particularly important because it offers a unique window into the structure of exotic nuclei near the proton dripline~\cite{Delion2006}. The two-proton decay is one of the major interest. It was expected to occur when one-proton emission is energetically prohibited as firstly predicted by Goldansky in 1960~\cite{Goldansky1960} and observed by experiment 20 years ago~\cite{Pfutzner2002}. Many discussions have been made about whether it is a true three-body decay (core+p+p) or a two-body decay (core+di-proton) and how it is related to the proton-proton correlation~\cite{Grigorenko2000, Grigorenko2009}. 

The $2p$ radioactivity has been observed not only in the ground states~\cite{Giovinazzo2002,Blank2005,Dossat2005,Mukha2007,Goigoux2016}, but also in various excited states~\cite{Egorova2012,Brown2015,Webb2019}. In recent years, systematic measurements have been made for the excited states of $^{10}$C~\cite{Charity2009,Charity2022}. The $1p$ and $2p$ decays have been observed for the $0^+_2$ and $2^+$ states, and the partial decay widths have been measured. There have been significant efforts to calculate the $2p$ decay. Some of them employ the $R$-matrix theory with the assumption of simultaneous emission of di-proton~\cite{Wigner1946,Lane1958,Descouvemont1989}. Combining with the shell model wave functions, the $2p$ decay of $^{12}$O, $^{18}$Ne and $^{45}$Fe have been studied~\cite{Barker1999,Barker2001,Brown2002,Brown2003}. The extensions to the three-body model were also made by many others~\cite{Grigorenko2009, Zhang2023, Wang2021}. However, the microscopic cluster models have not been applied to the study of $2p$ decay despite the importance of the cluster structure in light nuclei~\cite{Ikeda1968,Hoyle1954,Zhou2013}. Hence, developing the path of studying the proton decay with the microscopic cluster model is necessary.

In our previous works, a microscopic method to calculate the reduced width amplitude (RWA) has been developed~\cite{Chiba2017,Zhao2021,Taniguchi2021,Zhao2022}. Using this method, in the present work, we investigate proton decays from the excited states of $^{10}$C. We propose the prescriptions to determine the channel radius for the $R$-matrix calculation. The obtained results will be discussed in comparison with the recent experimental data~\cite{Charity2022}. We found that, the molecule-like cluster of $^{10}$C and $^{9}$B has a strong impact on the decay pattern of the $0^+_2$ and $2^+_2$ states of $^{10}$C.

This paper is organized as follows. In the next section, the theoretical framework to evaluate the $1p$ and $2p$ decays is briefly explained. In Sec.~\ref{sec:results}, we present the numerical results of the RWA and the determinations of the channel radius. Finally, we will discuss the results and make the summation.

\section{Theoretical Framework}
\label{sec:framework}

\subsection{The Hamiltonian and the wave function}
We combine the real-time evolution method (REM) with the generator coordinate method (GCM) to obtain the wave function of nuclei, in which the single-particle wave function $\phi(\bm{r},Z)$ is expressed in a Gaussian form multiplied by the spin-isospin part $\chi\tau$ as 
\begin{equation}
\begin{split}
\phi(\bm{r},Z) =
 (\frac{2\nu}{\pi})^{3/4}&\text{exp}[-\nu(\bm{r}-\frac{\bm{z}}{\sqrt{\nu}})^2+\frac{1}{2}\bm{z}^2]\chi\tau~,\\
 &Z\equiv(\bm{z}, a, b)~.
\end{split} 
\end{equation}
Here the coordinates $Z$ includes the spacial coordinates $\bm{z}$ and the spinor $a$ and $b$, $\chi = a\ket{\uparrow}+b\ket{\downarrow}$. The isospin part is $\tau=\{\text{proton or neutron}\}$. The harmonic oscillator parameter is set to $\nu=1/2b^2$ where $b=1.46$ fm, which reproduces the observed radius of $^4$He and is generally used as in Refs.~\cite{Itagaki2003,Furumoto2018} and our previous works~\cite{Zhao2021,Zhao2022}.

The wave function of the $\alpha$ cluster is constructed by the antisymmetrized wave function with $(0s)^4$ configuration as
\begin{equation}
\Phi_\alpha(\bm{z}_\alpha)=\mathcal{A}\{\phi_1(\bm{z}_\alpha,p_\uparrow)\phi_2(\bm{z}_\alpha,p_\downarrow)\phi_3(\bm{z}_\alpha,n_\uparrow)\phi_4(\bm{z}_\alpha,n_\downarrow)\}~, 
\end{equation}
by simply set the same spatial coordinates $\bm{z}_\alpha$ for four particles. 

The wave functions of nuclei composed of $\alpha$-clusters plus valence nucleons are represented by
\begin{equation}
\Phi(\bm{z}_{\alpha_1}...Z_1,Z_2...)=\mathcal{A}\{\Phi_\alpha(\bm{z}_{\alpha_1})...\phi(Z_1)\phi(Z_2)...\}~.
\end{equation}
With the help of the REM procedure introduced in Refs.~\cite{Zhou2020, Zhao2022}, many wave functions with different coordinates $Z$ are generated.  The total wave function is given as the superposition of the basis wave functions after the angular momentum projection,
\begin{equation}
\Psi^{J^\pi}_M = \sum_{i,K} f_{i,K} \hat{P}^{J^\pi}_{MK}\Phi_i~,
\end{equation}
where $\hat{P}^{J^\pi}_{MK}$ is the parity and the angular momentum projector. The coefficients of the superposition $f_{i,K}$ and the corresponding eigen-energy $E$ are obtained by solving the Hill-Wheeler equation. 

The Hamiltonian adopted in this work is given as
\begin{equation}
\hat{H}=\sum_{i=1}^A \hat{t}_i - \hat{T}_{c.m.} + \sum_{i<j}^A \hat{v}_N(\bm{r}_{ij}) + \sum_{i<j}^A \hat{v}_{C}(\bm{r}_{ij}) + \sum_{i<j}^A \hat{v}_{LS}(\bm{r}_{ij})~,
\end{equation}
where $\hat{t}_i$ and $\hat{T}_{c.m.}$ denote the kinetic energy operators of each nucleon and the center of mass, respectively. $\hat{v}_N$, $\hat{v}_C$, $\hat{v}_{LS}$ denote the effective central nucleon-nucleon interaction, the Coulomb interaction, and the spin-orbit interaction, respectively.

For the central nucleon-nucleon interaction, we use the Volkov No.2 interaction~\cite{Volkov1965}, which is expressed as
\begin{equation}
\begin{split}
\hat{v}_N(\bm{r}_{ij})=&(W - M\hat{P}^\sigma \hat{P}^\tau + B\hat{P}^\sigma - H\hat{P}^\tau)\\
&\times [V_1\text{exp}(-\bm{r}_{ij}^2/c_1^2)+V_2\text{exp}(-\bm{r}_{ij}^2/c_2^2)] ~,
\end{split}
\end{equation}
where $W$, $M$, $B$, and $H$ denote the Wigner, Majorana, Bartlett, and Heisenberg exchange parameters. The other parameters are, $V_1 = -60.65$ MeV, $V_2 = 61.14$ MeV, $c_1 = 1.80$ fm and $c_2 = 1.01$ fm. We use the G3RS potential~\cite{Tamagaki1968, Yamaguchi1979} as the spin-orbit interaction,
\begin{equation}
\hat{v}_{LS}(\bm{r}_{ij}) = V_{ls}(e^{-d_1 \bm{r}_{ij}^2} - e^{-d_2 \bm{r}_{ij}^2})\hat{P}_{31}\hat{L}\cdot \hat{S}~.
\end{equation}
Here $\hat{P}_{31}$ projects the two-body system into the triplet-odd state, which can be expressed as $\hat{P}_{31}=\frac{1+\hat{P}^\sigma}{2}\cdot\frac{1+\hat{P}^\tau}{2}$. The Gaussian range parameters $d_1$ and $d_2$ are set to be $5.0$ fm$^{-2}$ and $2.778$ fm$^{-2}$, respectively. In this work, the exchange parameters of central interaction and the strength of the spin-orbit interaction are slightly modified to reproduce the decay $Q$-values. The determination of these parameters will be explained in the next section in more detail.

\subsection{Reduced width amplitude and decay width}

The reduced width amplitude (RWA) can be regarded as the wave function of a cluster in a nucleus, which can be used for the calculation of many other quantities through the $R$-matrix theory~\cite{Descouvemont2010}. It is defined as the overlap amplitude between the $A$-body wave function of the mother nucleus $\Psi$ and the decay channel composed of the residue nuclei with mass numbers $A_1$ and $A_2$,
\begin{equation}
\label{eq:rwa}
ay_l(a) = a\sqrt{\frac{A!}{(1+\delta_{A_1A_2})A_1!A_2!}}\langle \frac{\delta(r-a)}{r^2}\Psi_{A_1}\Psi_{A_2}Y_l(\hat{r})|\Psi\rangle~.
\end{equation}
Here $\Psi_{A_1}$ and $\Psi_{A_2}$ are the wave functions of the two residues and $l$ represents the relative angular momentum between them. Eq.~\ref{eq:rwa} is calculated by using the Laplace expansion method~\cite{Chiba2017}. In this work, the wave functions of the mother and daughter nuclei are calculated by the GCM explained above. The two-proton wave function is approximated by a single Slater determinant projected to $J^{\pi}=0^+$
\begin{equation}
\begin{split}
&\Psi^{J^\pi=0^+} = \sum_{K} \hat{P}^{J^\pi}_{MK}\Phi_{2p}~,\\
&\Phi_{2p}=\mathcal{A}\{\phi_1(d/2,p_\uparrow)\phi_2(-d/2,p_\downarrow)\}~, 
\end{split}
\end{equation}
where $d$ denotes the distance between two protons which is set to be $d=0.5$ fm describing a compact diproton state. Hence, we estimate the $2p$ decay width by assuming the two-body decay of $^{10}$C($0^+_2$)$\rightarrow$$^8$Be($0^+$)+$2p$.

According to the $R$-matrix theory, the partial width for the two-body decay $A \rightarrow A_1+A_2$ is given by the square of the RWA and the reduced mass $\mu$ as
\begin{equation}
\label{eq:width}
\Gamma_{A_1+A_2}(a)=P_{l,\eta}(a)\frac{\hbar^2}{2\mu a}|ay_l(a)|^2~.
\end{equation}
In this equation, $P_{l,\eta}(a)$ is the penetrability factor given as
\begin{equation}
P_{l,\eta}(a)=\frac{2ka}{F_{l,\eta}(ka)^2+G_{l,\eta}(ka)^2}~,
\end{equation}
where $F_{l,\eta}$ and $G_{l,\eta}$ are the regular and irregular Coulomb functions. The wave number $k$ and the dimensionless Sommerfeld parameter $\eta$ are defined by the decay $Q$-value and the reduced mass as follows.
\begin{equation}
\begin{split}
&k=\sqrt{2\mu Q/\hbar^2}\\
&\eta=Z_1Z_2e^2\mu/2\pi\epsilon_0\hbar^2k~.
\end{split}
\end{equation}

From Eq.~\ref{eq:width}, the partial width is obtained as a function of the channel radius $a$ between two nuclei. Theoretically, the channel radius is the position where the nuclear force between the decay residues and the decay particle is negligible. Therefore, it should be chosen as the point where the RWA is smoothly connected to the Coulomb function.

\section{Results}
\label{sec:results}
\subsection{Decay of the $0^+_2$ and $2^+_2$ states of $^{10}$C}
The experiments~\cite{Charity2022} showed that the $^{10}$C($0^+_2$) predominantly decays to the ground state of $^8$Be by the $2p$ emission and the $1p$ decay is suppressed, although both decay channels are open. This surprising result was explained by the mismatch of the valence proton configurations in $^{10}$C($0^+_2$) and $^9$B($3/2^-$) within the context of the shell model~\cite{Fortune2006}. Alternatively, the molecular orbit model might be able to give a more reasonable explanation for this structure mismatch as the $^{10}$C($0^+_2$) has a pronounced cluster structure.  In the molecular orbit model, $^{10}$C, which is the mirror nucleus of $^{10}$Be, is modeled as two $\alpha$ clusters coupled with two valence protons occupying so-called molecular orbits in analogy with the atomic molecules. It has been discussed that the valence protons occupy the $\pi$-orbit (negative-parity and mainly composed of $p$-shell) in the ground, $2^+_1$ and $2^+_2$ states, whereas they occupy the $\sigma$-orbit (positive-parity, mainly composed of $sd$-shell) in the $0^+_2$ state. This structural difference of the $0^+_2$ and $2^+_2$ states impacts their decay pathways.

First, let us consider the decay of the $^{10}$C($0^+_2$) state. After the emission of $1p$ occupying the $\sigma$-orbit, the residual nucleus $^{9}$B also has a proton in the $\sigma$-orbit, whose wave function largely overlaps with the first excited state ($1/2^+_1$) of $^{9}$B, but almost orthogonal to the ground state. Therefore, the $1p$ decay to the ground state of $^{9}$B should be suppressed due to the structural mismatch. Note that the energy of $^{9}$B($1/2^+$), which is a broad resonance, is higher than that of $^{10}$C($0^+_2$), and hence, the decay to the $^{9}$B($1/2^+$) may also be suppressed. Consequently, the decay to the $^{8}$Be by the simultaneous $2p$ emission might be the dominant decay pathway. The situation is quite different for the $^{10}$C($2^+_2$) state. After the $1p$ emission, the wave function of the residual nucleus largely overlaps with the $^{9}$B ground state. Therefore the $1p$ decay should be dominant. This argument is based on qualitative expectations and has not been quantitatively confirmed by the nuclear structure model calculations. Therefore, in this study, we examine it numerically by using a microscopic cluster model that can properly describe the molecular structure of $^{10}$C and $^{9}$B.

\subsection{Decay $Q$-values and interaction parameters}
We first determine the parameters of the central and spin-orbit interactions to reproduce the experimental decay $Q$-values. The $0^+_2$ state of $^{10}$C has two decay pathways: one is to $^9$B($3/2^-$) via emission of a single proton with the decay $Q$-value of $1.21$ MeV, and the other is to $^8$Be($0^+$) via emission of two protons with the $Q$-value of $1.40$ MeV. Despite the negative $Q$-value, the decay to the $^9$B($1/2^+$) might be also allowed, because it is a broad resonance. However, we will not investigate this pathway in the present paper. We also calculate the decay pathways of $^{10}$C($2^+_2$)$\rightarrow$$^9$B($3/2^-$)+$p$ with the $Q$-value of $1.37$ MeV. 

\begin{figure*}[htbp]
 \begin{center}
  \includegraphics[width=0.7\hsize]{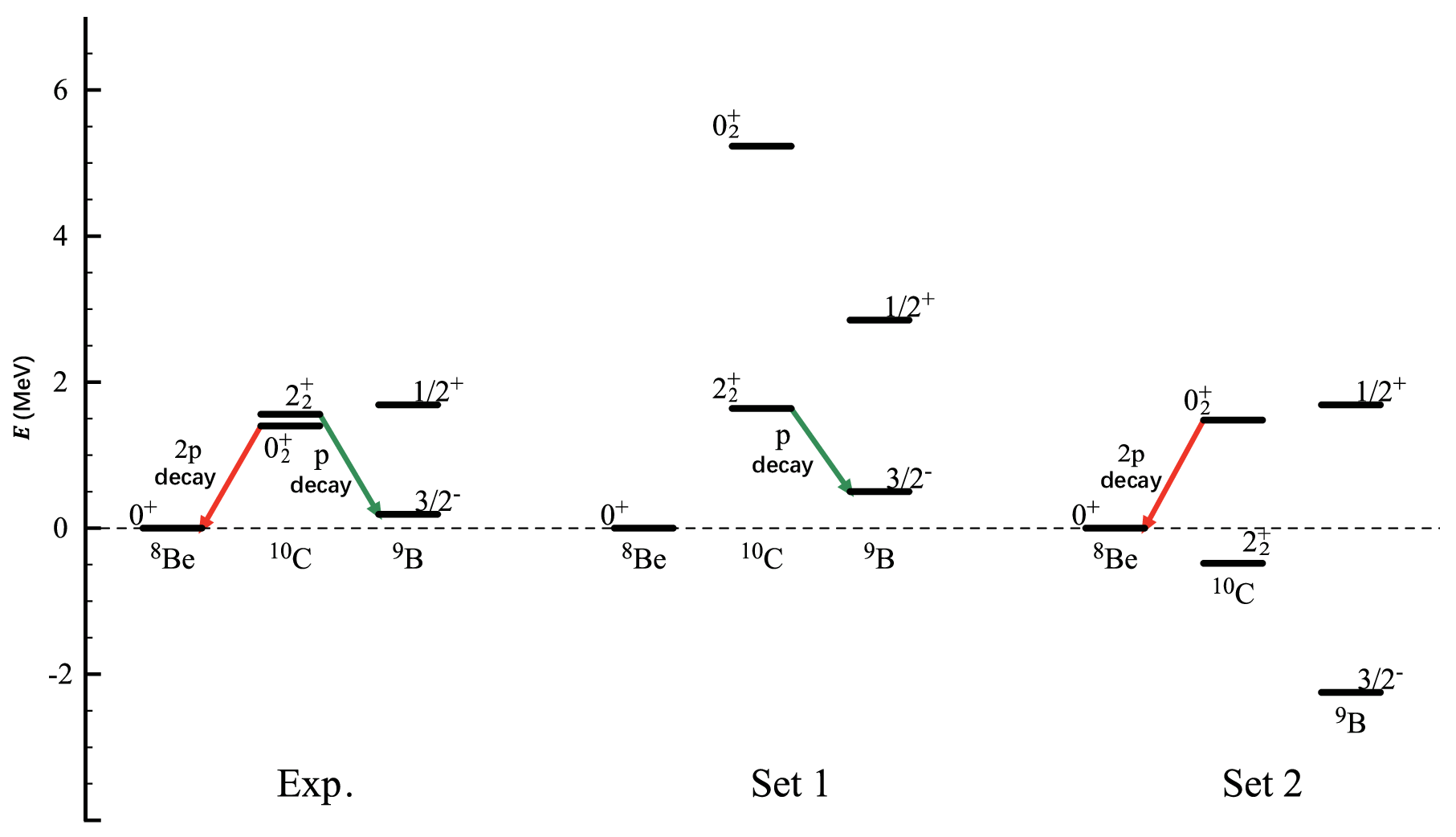}
  \caption{\label{fig:ene}Energy spectra measured from the energy of $^{8}$Be($0^+$). ``Expt." denotes the experimental data~\cite{NNDC}. The $5.22$ MeV excited state in the experiment is denoted as the $0^+_2$ state of $^{10}$C. ``Set 1" and ``Set 2" denote the results calculated using two sets of interaction parameters (see text). The arrow lines are the results adopted for the following discussions.} 
  \end{center}
\end{figure*}

An ordinary set of the parameters, which we call set 1 in the following, is $W=0.4$, $M=0.6$, $B=H=0.125$, and $V_{ls}=2000$ MeV. The set 1 has been widely used in previous calculations~\cite{Zhao2022,Kanada2012,Itagaki2003}. However, it cannot reproduce the $Q$-values as shown in Fig.~\ref{fig:ene}.
Hence, we introduce a slightly modified parameter set denoted by set 2 that is $W=0.44$, $M=0.56$, $B=H=0.2$, and $V_{ls}=2400$ MeV, which reproduces the $Q$-values for the $^{10}$C($0^+_2$)$\rightarrow$$^8$Be($0^+$)+$2p$. The $Q$-value for $1p$ decay of $^{10}$C($0^+_2$) still cannot be reproduced, but it will not strongly affect the calculating result as we will see later. We also calculate the $^{10}$C($2^+_2$)$\rightarrow$$^9$B($3/2^-$)+$p$ channel. For this case, set 1 is more appropriate. In short, we use set 2 for all calculations, except for the $1p$ decay of the $2^+_2$ state.

\subsection{RWA and decay width}
\begin{figure}[htbp]
\begin{center}
  \includegraphics[width=1.0\hsize]{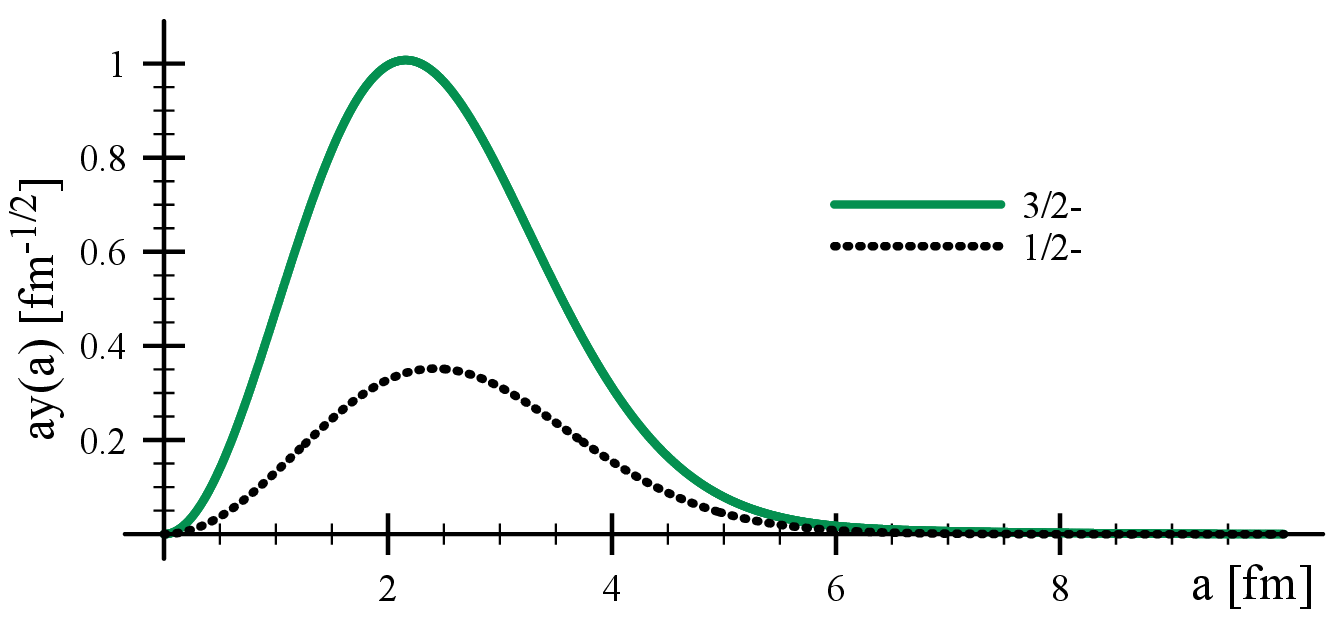}
  \caption{\label{fig:lr}The RWAs for the decays of $^{10}$C($2^+_2$)$\rightarrow$$^9$B($3/2^-$)+$p$ with different angular momentum in $z$ direction of the proton.} 
  \end{center}
\end{figure}

In the decay channel of $^{10}$C($2^+_2$)$\rightarrow$$^9$B($3/2^-$)+$p$, several relative angular momenta between $^9$B and the proton are allowed, i.e. $J^{\pi}=1/2^-$, $3/2^-$, $5/2^-$, and $7/2^-$. In Fig.~\ref{fig:lr}, we compare the RWAs for the $J^{\pi}=1/2^-$ and $3/2^-$ channels.
It shows that the contribution of the $J^{\pi}=1/2^-$ channel is small, and we found that its partial width is only a few keV. We also confirmed that the other angular momenta, $5/2^-$, and $7/2^-$ are negligible. Therefore, in the present discussions, we only consider the $J^{\pi}=3/2^-$ case for the $^{10}$C($2^+_2$)$\rightarrow$$^9$B($3/2^-$)+$p$ channel.

\begin{figure}[htbp]
\begin{center}
  \includegraphics[width=1.0\hsize]{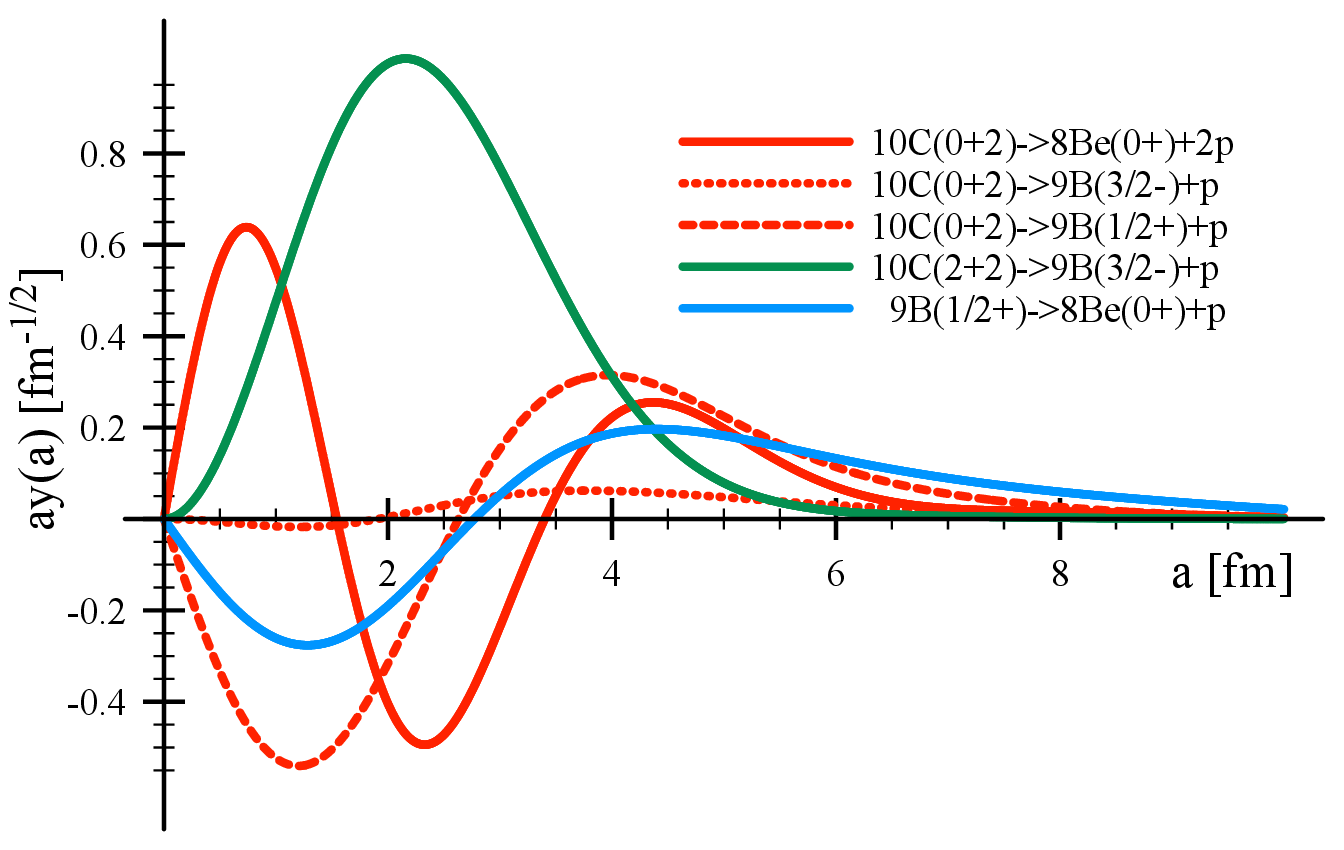}
  \caption{\label{fig:rwa}The RWAs for the decays of the $^{10}$C($0^+_2$), $^{10}$C($2^+_2$) and $^9$B($1/2^+$) states to the ground and first excited $1/2^+$ states of $^9$B.} 
  \end{center}
\end{figure}

Fig.~\ref{fig:rwa} shows the RWAs with respect to the decays of the $^{10}$C($0^+_2$) and $^{10}$C($2^+_2$) to the ground ($3/2^-$) and first excited ($1/2^+$) states of $^9$B as well as the decay of the $^9$B($1/2^+$) to $^8$Be($0^+$).
Since the RWA is the overlap of the wave functions between the mother nucleus and the decay residues, the amplitude of the RWA reflects the structural similarity between them. As already explained, $^{10}$C($0^+_2$) has similar structure with the $^9$B($1/2^+$) but not $^9$B($3/2^-$). Consequently, the amplitude of the RWA for the $^{10}$C($2^+_2$)$\rightarrow$$^9$B($3/2^-$)+$p$ and $^{10}$C($0^+_2$)$\rightarrow$$^8$Be($0^+$)+$2p$ channels are large, whereas that of the $^{10}$C($0^+_2$)$\rightarrow$$^9$B($3/2^-$)+$p$ channel is negligible. These indicate that the $^{10}$C($0^+_2$) decays to $^8$Be($0^+$) via $2p$ emission, or $1p$ emission to $^9$B($1/2^+$). Furthermore, because of the negative $Q$-value of the decay to $^9$B($1/2^+$), the $2p$ emission becomes the main decay pathway of $^{10}$C($0^+_2$). Contrary, $^{10}$C($2^+_2$) has the similar structure to the $^9$B($3/2^-$), which makes the amplitude of the RWA for the $^{10}$C($2^+_2$)$\rightarrow$$^9$B($3/2^-$)+$p$ channel large. Similarly to the decay of $^{10}$C($0^+_2$)$\rightarrow$$^8$Be($0^+$)+$2p$, the valence proton in $^{9}$B($1/2^+$) occupies the same orbit as in $^{10}$C($0^+_2$), so that the amplitude of $^{9}$B($1/2^+$)$\rightarrow$$^8$Be($0^+$)+$p$ is large too. The longer tail of the RWA for $^{9}$B($1/2^+$)$\rightarrow$$^8$Be($0^+$)+$p$ decay channel indicates the broad width of $^9$B($1/2^+$).

\begin{figure}[htbp]
\begin{center}
  \includegraphics[width=1.0\hsize]{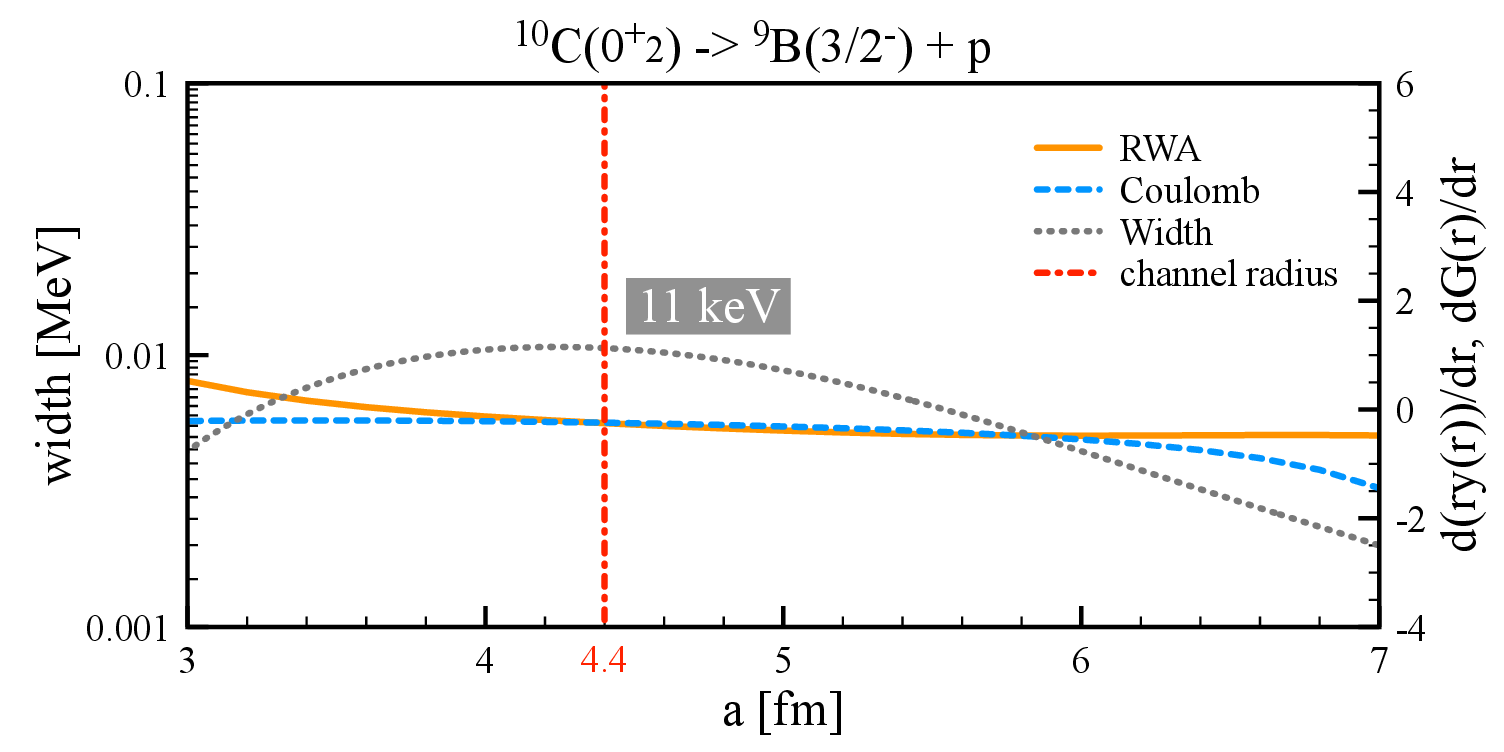}
  \includegraphics[width=1.0\hsize]{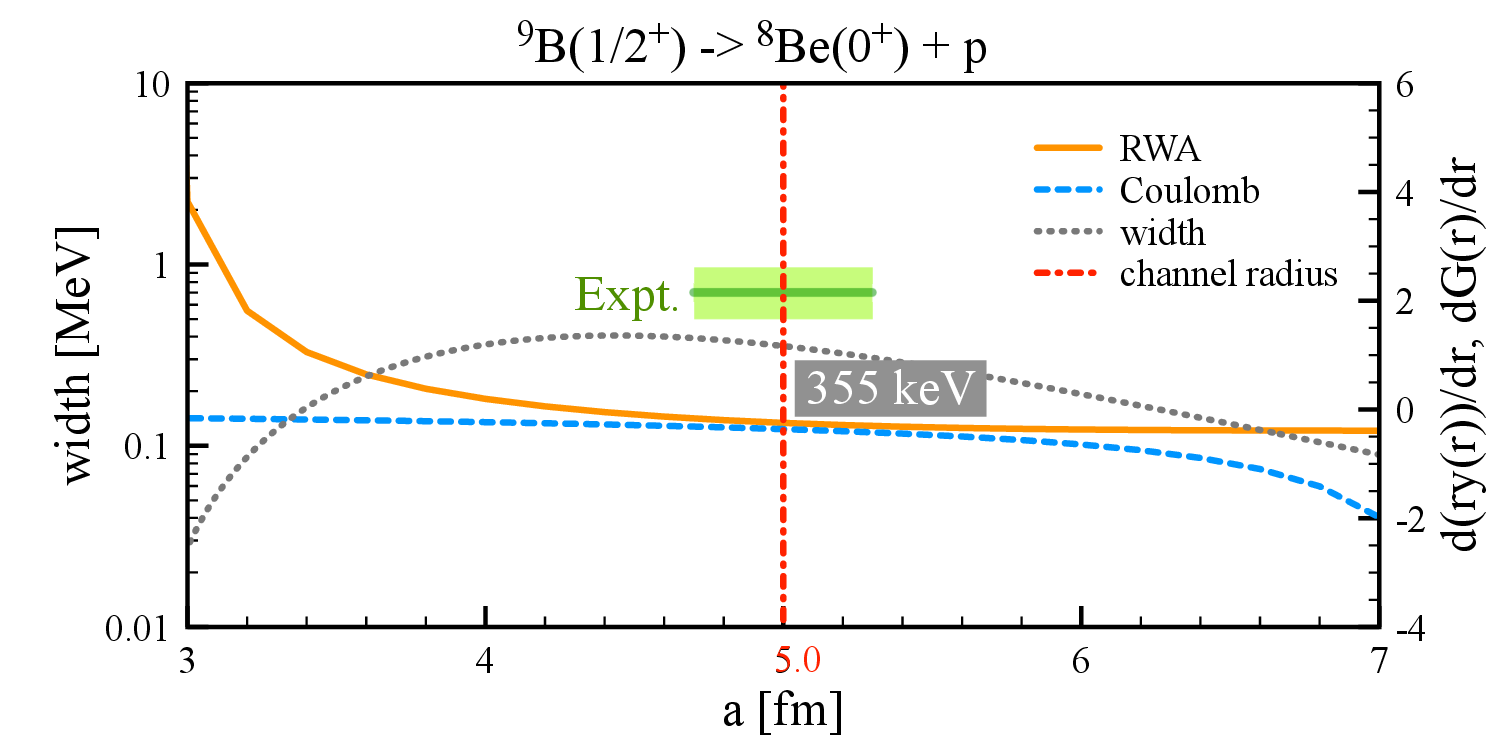}
  \caption{\label{fig:widthC}The calculated width (left scale) and the comparison between the RWA and Coulomb function (right scale) along with the distance $a$ for the $^{10}$C($0^+_2$)$\rightarrow$$^9$B($3/2^-$)+$p$ channel (upper figure) and $^{9}$B($1/2^+$)$\rightarrow$$^8$Be($0^+$)+$p$ channel (lower figure), respectively. The red dash-dotted line denotes the choice of the channel radius. The result shown in gray box is the corresponding width. The green box denotes the experimental data with the error bar.} 
  \end{center}
\end{figure}

With the obtained RWA, we calculate the partial width as a function of the channel radius $a$ as shown in Fig.~\ref{fig:widthC}. The channel radius should be determined so that the RWA has the same asymptotic as the Coulomb function. For this purpose, Fig.~\ref{fig:widthC} also compares the logarithmic derivatives of the RWA and the Coulomb function ( see Appendix.~\ref{appx}).
In the figures, the orange line and the blue dashed line are the derivatives of the RWA and Coulomb function, respectively. At the channel radius, these two lines should be identical. For example, these two lines are almost identical around $a \approx 5$ fm for the $^{10}$C($0^+_2$)$\rightarrow$$^9$B($3/2^-$)+$p$ and $^{9}$B($1/2^+$)$\rightarrow$$^8$Be($0^+$)+$p$ decays. For the $^{10}$C($0^+_2$)$\rightarrow$$^9$B($3/2^-$)+$p$ channel, the upper figure, we have chosen the channel radius to be about $4.4$ fm, while the $^{9}$B($1/2^+$)$\rightarrow$$^8$Be($0^+$)+$p$ channel, it is about $5.0$ fm.

\begin{figure}[htbp]
\begin{center}
  \includegraphics[width=1.0\hsize]{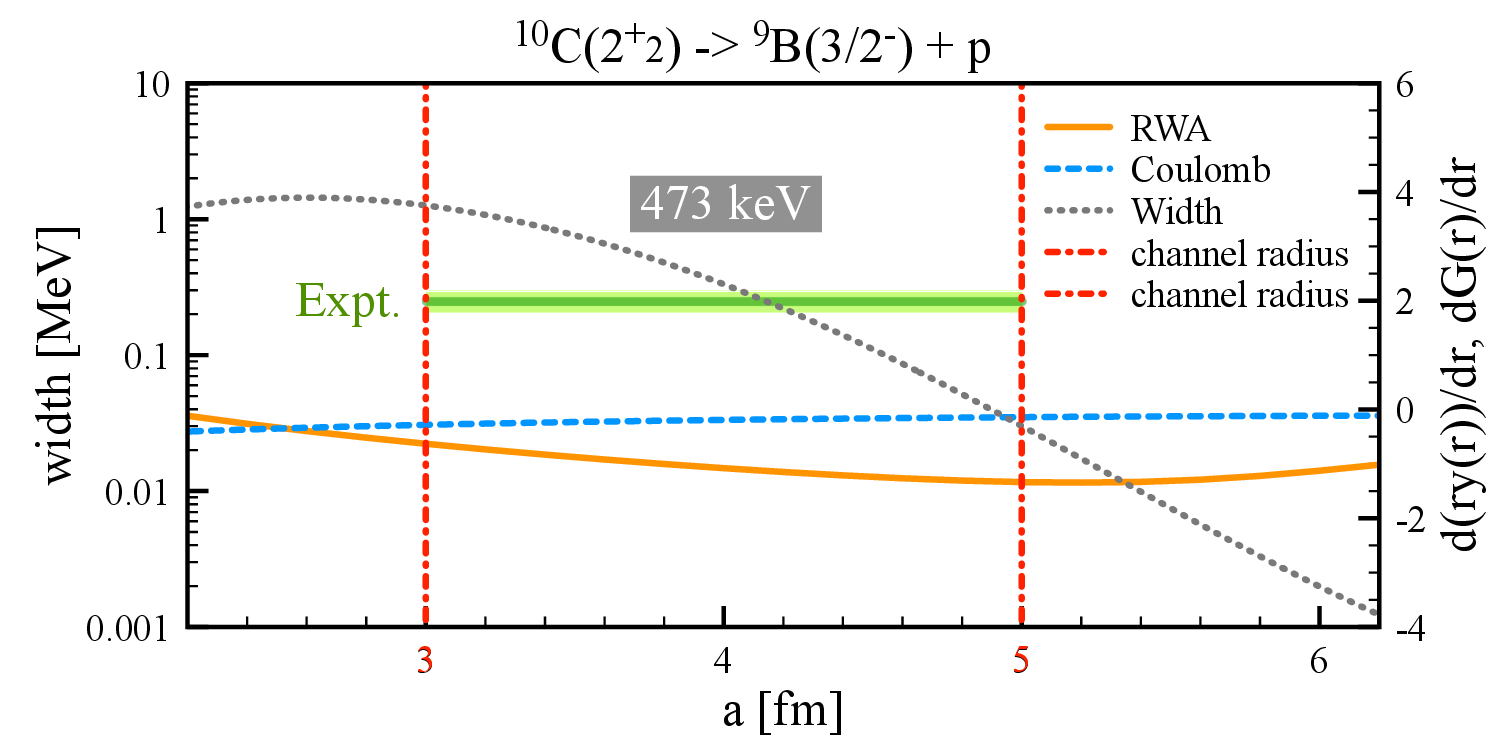}
  \includegraphics[width=1.0\hsize]{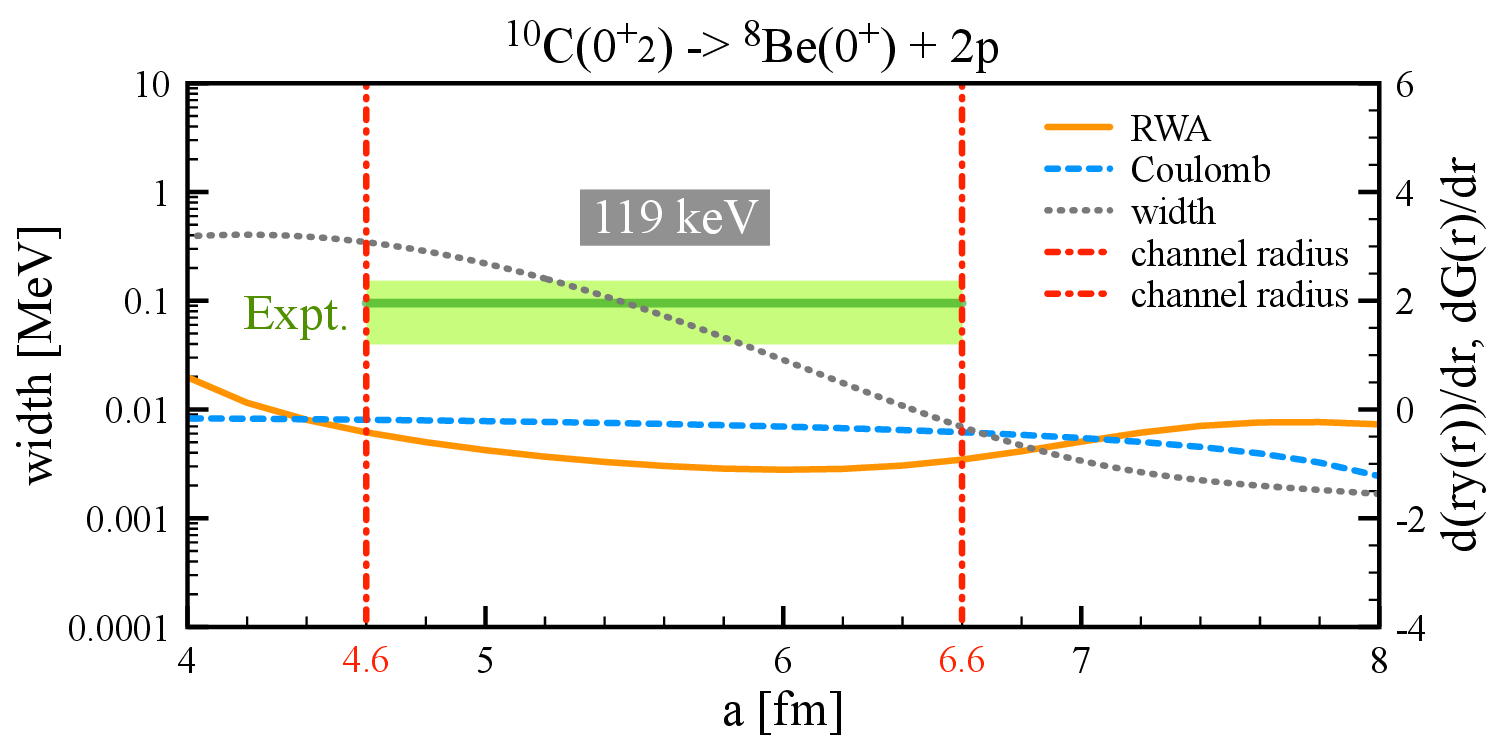}
  \caption{\label{fig:widthCa}Same as previous figure but for the $^{10}$C($2^+_2$)$\rightarrow$$^9$B($3/2^-$)+$p$ channel (upper figure) and $^{10}$C($0^+_2$)$\rightarrow$$^8$Be($0^+$)+$2p$ (lower figure), respectively. The channel radii indicate the upper and lower limit for the width result. The average value in this region is derived as the width result shown in the gray box.} 
  \end{center}
\end{figure}

Because of the limitation in accuracy on the calculation of RWA, the obtained results do not always follow the Coulomb function well. In Fig.~\ref{fig:widthCa}, we show the calculated width and the comparison between the RWA and Coulomb function for the $^{10}$C($2^+_2$)$\rightarrow$$^9$B($3/2^-$)+$p$ and $^{10}$C($0^+_2$)$\rightarrow$$^8$Be($0^+$)+$2p$ channels.
For these cases, we cannot determine the channel radius, and hence, we estimate an average of the expected decay widths. Comparing the RWA and Coulomb function, the starting point is taken to be the point behind where they first contact. The range of 2 fm after the start point is taken as the selecting region. For $^{10}$C($2^+_2$)$\rightarrow$$^9$B($3/2^-$)+$p$ channel, we take the region as $3\sim 5$ fm. For $^{10}$C($0^+_2$)$\rightarrow$$^8$Be($0^+$)+$2p$ channel, we take the region as $4.6\sim 6.6$ fm. 

After determining the channel radius, we obtain the partial widths of the $1p$ and $2p$ decays. The widths obtained in the present work are compared with the experimental data in Table.~\ref{table:width}.
\begin{table}[htbp]
  \begin{center}
    \caption{The calculating width results obtained in this work and the corresponding experimental data.} \label{table:width}
    \vspace{2mm}
 \begin{tabular*}{7cm}{ @{\extracolsep{\fill}} l l c c}
    \hline
                                &Branch       &$\Gamma$[keV]      &Expt.[keV]\\
    \hline
$^{10}$C($0^+_2$) &$^9$B$_{g.s.}+p$      &11       &$<$10~\cite{Charity2022} \\
                                &$^8$Be$_{g.s.}+2p$   &119      &96(57)~\cite{NNDC} \\
$^{10}$C($2^+_2$) &$^9$B$_{g.s.}+p$      &473      &250(46)~\cite{Charity2022} \\
$^{9}$B($1/2^+$)     &$^8$Be$_{g.s.}+p$    &355      &$700(^{+270}_{-200})$~\cite{Scholl2011} \\
   \hline
  \end{tabular*}
  \end{center}
\end{table} 
All of the results agree with the recent experimental data in order of magnitude. We can see that the $1p$ decay is much suppressed in $^{10}$C($0^+_2$), which is only about $11$ keV. This result is consistent with the previous theoretical predictions~\cite{Fortune2006,Arai1996,Tanaka1999}. We also show the decay width of $^9$B($1/2^+$), which slightly underestimates the experimental data. It might be because it is a broad resonance state and its wave function is difficult to calculate.

What should be noted in the end is that such consistent $2p$ decay width of $^{10}$C($0^+_2$) is obtained under the assumption of a compact diproton structure for the wave function of two protons. It means that the two-body model can be a proper assumption for the $2p$ decay in $^{10}$C($0^+_2$). However, we still should be careful to apply the two-body model on other $2p$ decay, for example in $^6$Be, as they may follow the three-body decay model as discussed in Ref.~\cite{Grigorenko2009}.

\section{Summary}
\label{sec:summary}
During this work, we perform fully microscopic calculations with the cluster model to calculate the RWA and width for the proton decay channel. The way of determining the channel radius has been proposed during the calculations. The $1p$ decay in $^{10}$C($0^+_2$), $^{10}$C($2^+_2$), and $^{9}$B($1/2^+$) have been calculated. Within the diproton assumption, the $2p$ decay in $^{10}$C($0^+_2$) is also calculated. The width results are in good agreement with the recent experimental data in order of magnitude. These results demonstrate that the suppression of $1p$ decay in $^{10}$C($0^+_2$) is due to the mismatch structure of the valence protons, and the $2p$ decay in this state can be well explained by the two-body decay model. This work is the first time to apply the microscopic cluster model calculation to obtain the partial decay width. The calculation procedure still needs to be further developed, for example by applying it to the three-body decay model.

% cSpell: disable 
\begin{acknowledgments}
The authors thank Dr. Zaihong Yang for the fruitful discussions. This work was supported by National Natural Science Foundation of China [Grant Nos. 12175042, 12275081, 12275082, 12305123], and JSPS KAKENHI [Grant Nos. 19K03859, 21H00113 and 22H01214]. Numerical calculations were performed in the Cluster-Computing Center of School of Science (C3S2) at Huzhou University.
\end{acknowledgments}

\begin{appendix}
\section{Coulomb function}
\label{appx}
The solution of the following equation:
\begin{equation}
\frac{d^2w}{dz^2}+(1-\frac{2\eta}{z}-\frac{l(l+1)}{z^2})w=0
\end{equation}
which has parameter $\eta$ is given by two linearly independent solutions with arbitrary constants $C_1$ and $C_2$ as
\begin{equation}
w(z)=C_1F_l(\eta,z)+C_2G_l(\eta,z).
\end{equation}
Here $F_l(\eta,z)$ is the regular Coulomb function and $G_l(\eta,z)$ is the irregular Coulomb function.

In the region where only the Coulomb potential is present, the Coulomb potential is $\frac{1}{4\pi\epsilon_0}\frac{Z_1Z_2e^2}{r}$ and then the Schr{\"o}dinger equation becomes
\begin{equation}
\frac{d^2u(r)}{dr^2}+(\frac{2mE}{\hbar^2}-\frac{Z_1Z_2e^2m}{2\pi\epsilon_0\hbar^2}\frac{1}{r}-\frac{l(l+1)}{r^2})u(r)=0
\end{equation}
In the case of unbound state ($E>0$), we can define $k=\sqrt{2mE/\hbar^2}$ and $\eta=Z_1Z_2e^2m/2\pi\epsilon_0\hbar^2k$. Then we can obtain
\begin{equation}
\begin{split}
\frac{d^2u(r)}{dr^2}+(k^2-\eta k\frac{1}{r}-\frac{l(l+1)}{r^2})u(r)&=0\\
\frac{d^2u(r)}{k^2dr^2}+(1-\frac{\eta}{2}\frac{2}{kr}-\frac{l(l+1)}{k^2r^2})u(r)&=0
\end{split}
\end{equation}
Therefore, the solution of this equation is the Coulomb functions
\begin{equation}
u(kr)=C_1F_l(\eta/2,kr)+C_2G_l(\eta/2,kr).
\end{equation}
For the physical meaning, the wave function can only satisfy the Coulomb $G$ function with a constant $C$ as
\begin{equation}
u(kr)=CG_l(\eta/2,kr),
\end{equation}
where the wave function $u(kr)$ is just the RWA $ry_l(r)$ calculated from the framework.
By comparing logarithmic derivatives of these functions as following equations:
\begin{equation}
\label{eq:ratiow}
\frac{d}{dr}\ln{(ry_l(r))} = \frac{[ry_l(r)]'}{ry_l(r)}~,~~\frac{d}{dr}\ln{G(r)} =\frac{G'(r)}{G(r)}~,
\end{equation}
the position of where they just connect can be determined.

\end{appendix}

\end{document}